\documentclass[letterpaper,twocolumn,prb]{revtex4-1}

\usepackage[dvipdfmx]{graphicx}
\usepackage{bm}
\usepackage{lipsum}
\usepackage{color}
\usepackage[justification=raggedright]{caption}

\begin{document}

\title{Pressure suppression of the excitonic insulator state 
in Ta$_2$NiSe$_5$ observed by\\
optical conductivity}
\author{H.~Okamura}\altaffiliation[Electronic address: ]{ho@tokushima-u.ac.jp}
\author{T. Mizokawa$^1$}
\author{K. Miki}
\author{Y. Matsui}
\author{N. Noguchi}
\author{N. Katayama$^2$}
\author{H. Sawa$^2$}
\author{M. Nohara$^3$}
\author{Y. Lu$^4$}
\author{H. Takagi$^{5,6}$}
\author{Y. Ikemoto$^7$}
\author{T. Moriwaki$^7$}
\affiliation{Department of Applied Chemistry, 
Tokushima University, Tokushima 770-8506, Japan}
\affiliation{$^1$Department of Applied Physics, Waseda University, 
Tokyo 169-8555, Japan}
\affiliation{$^2$Department of Applied Physics, Nagoya University, 
Nagoya 464-8603, Japan}
\affiliation{$^3$Department of Quantum Matter, Hiroshima University, 
Higashi-Hiroshima 739-8503, Japan}
\affiliation{$^4$College of Materials Science and Engineering, 
National Engineering Research Center for Magnesium Alloys, 
Chongqing University, Chongqing 400044, China}
\affiliation{$^5$Department of Physics, University of Tokyo, 
Tokyo 113-0013, Japan}
\affiliation{$^6$Max Planck Institute for Solid State Research, 
70569 Stuttgart, Germany}
\affiliation{$^7$Japan Synchrotron Radiation Research Institute, 
Sayo 679-5198, Japan}

\date{\today}

\begin{abstract}
The layered chalcogenide Ta$_2$NiSe$_5$ has recently 
attracted much interest as a strong candidate for the 
long sought excitonic insulator (EI).   
Since the physical properties of an EI are expected 
to depend sensitively on the external pressure, 
it is important to clarify the pressure evolution of 
microscopic electronic state in Ta$_2$NiSe$_5$.   
Here we report the optical conductivity [$\sigma(\omega)$] 
of Ta$_2$NiSe$_5$ measured at high pressures to 10~GPa 
and at low temperatures to 8~K.    
With cooling at ambient pressure, $\sigma(\omega)$ develops 
an energy gap of about 0.17~eV and a pronounced excitonic 
peak at 0.38~eV, as already reported in the literature.  
Upon increasing pressure, the energy gap becomes narrower 
and the excitonic peak is broadened.  Above a structural 
transition at $P_s \simeq$ 3~GPa, the energy gap becomes 
partially filled, indicating that Ta$_2$NiSe$_5$ is a 
semimetal after the EI state is suppressed by pressure.   
At higher pressures, $\sigma(\omega)$ exhibits metallic 
characteristics with no energy gap.   
The detailed pressure evolution of $\sigma(\omega)$ is 
presented, and 
discussed mainly in terms of a weakening of excitonic 
correlation with pressure.  
\end{abstract}

\maketitle
\section{Introduction}
Excitonic insulator (EI) is an unconventional insulator 
in which the attractive correlation between electrons and 
holes results in a collective condensation of electron-hole 
($e$-$h$) pairs (excitons) and an energy gap at the Fermi 
level ($E_{\rm F}$).  EI was first predicted theoretically in 
the 1960s,\cite{EI-review} with a phase diagram 
schematically shown in Fig.~1.  
%
\begin{figure}[b]
\begin{center}
\includegraphics[width=0.4\textwidth]{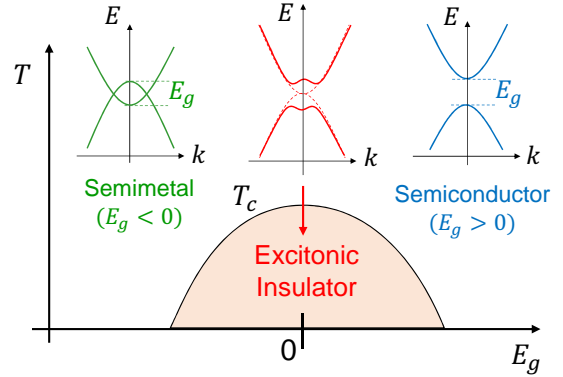}
\caption{Schematic phase diagram of EI in terms 
of the temperature ($T$) and the one-particle 
band gap in the absence of $e$-$h$ correlation 
($E_g$).  
Positive and negative values of $E_g$ correspond to 
semiconducting and semimetallic states, respectively.   
An EI state is expected in the vicinity of $E_g$=0 
below the transition temperature ($T_c$), 
with a characteristic flattening of the band 
edges as indicated.  
}
\end{center}
\end{figure}
The starting material for EI can be either a semimetal 
($E_g <$ 0) or semiconductor ($E_g >$ 0), where 
$E_g$ indicates the one-particle band gap in the 
absence of $e$-$h$ correlation.  An EI state is expected 
in the vicinity of $E_g$=0 below the transition 
temperature $T_c$ if the exciton binding energy $E_b$ 
is larger than $|E_g|$.  
Another feature of the EI is a characteristic 
flattening of the bands as indicated in Fig.~1.

Despite the theoretical interest on EI, only a few 
compounds had been considered as candidates for EI, 
including TmSe$_{1-x}$Te$_x$\cite{wachter} and 
1$T$-TiSe$_2$.\cite{TiSe2}  
In 2009, Wakisaka {\it et al.}\cite{wakisaka} suggested 
that Ta$_2$NiSe$_5$ should be an EI on the basis of 
angle-resolved photoemission spectroscopy (ARPES) data.  
Ta$_2$NiSe$_5$ has a layered crystal structure, and exhibits 
a structural phase transition at $T_c$=328~K, where the crystal 
symmetry changes from orthorhombic to monoclinic with 
decreasing temperature ($T$).\cite{sunshine,salvo}  
Below $T_c$, the resistivity increases rapidly with cooling, 
indicating an energy gap of about 0.2~eV.\cite{salvo} 
The ARPES study at low $T$ found an unusually flat 
dispersion at the top of valence band, which was ragarded 
as strong evidence for an EI.\cite{wakisaka}   
Further ARPES study\cite{seki} and band 
calculation\cite{seki,kaneko} 
indicated that the flat band feature persisted even 
above $T_c$, and suggested the low-$T$ phase of 
Ta$_2$NiSe$_5$ to be an EI in the strong coupling 
regime caused by a condensation of preformed 
excitons.\cite{kaneko,seki}   
It was pointed out that above $T_c$ the hybridization 
between the Ta 5d conduction band and Ni 3d valence 
band at $\Gamma$ point should be forbidden by symmetry.  
It was also suggested that the structural change below 
$T_c$ was mainly driven by the exciton condensation, 
rather than by a structural instability due to 
electron-lattice ($e$-$l$) coupling.\cite{kaneko}  
Lu et al\cite{lu} reported transport and optical data 
of Ta$_2$Ni(Se,S)$_5$ and Ta$_2$Ni(Se,Te)$_5$, where 
the chemical substitution and external pressure were 
used to control $E_g$ of Ta$_2$NiSe$_5$.  
The optical conductivity [$\sigma(\omega)$] 
of Ta$_2$NiSe$_5$ at low $T$ clearly exhibited a 
pronounced peak of excitonic origin at about 0.4~eV, 
and an energy gap of about 0.16~eV.   An experimental 
phase diagram of Ta$_2$NiSe$_5$ was constructed from 
the transport data, which was indeed consistent with 
the predicted one depicted in Fig.~1 and strongly 
suggested Ta$_2$NiSe$_5$ to be an EI.\cite{lu}  
Detailed $\sigma(\omega)$ data\cite{boris} were 
analyzed by a theoretical study,\cite{sugimoto} 
which again suggested that Ta$_2$NiSe$_5$ should 
be an EI in the strong coupling regime below $T_c$, 
and it should be a semiconductor ($E_g >$ 0) 
above $T_c$.\cite{sugimoto}

Many more works on Ta$_2$NiSe$_5$ have also been 
reported, including 
ARPES,\cite{mor1,okazaki1,mizokawa,tang,okazaki2,saha,
fukutani1,chen,watson,fukutani2} 
optical conductivity,\cite{seo,matsubayashi} 
Raman scattering,\cite{yan,raman-MPI,sood,kim,volkov} 
ultrafast laser spectroscopy,\cite{weudehausen,mor2,okamoto} 
resonant inelastic X-ray scattering,\cite{rixs1,rixs2} 
high pressure studies,\cite{matsubayashi,sood,sawa} 
scanning tunneling spectroscopy,\cite{STM} 
and theoretical analyses.\cite{watson,subedi,mazza}   
While many of these works support that the transition 
at $T_c$ is driven by exciton condensation, some of them 
question such scenario.\cite{watson,rixs2,subedi}  
For example, a semimetallic ($E_g <$ 0) band dispersion 
around $E_{\rm F}$ has been reported by an ARPES study 
above $T_c$.\cite{watson}  
Theoretically, it has been shown that a flat valence 
band may be obtained by a band calculation without $e$-$h$ 
correlation,\cite{watson} and that there can be a sizable 
($\sim$ 0.1~eV) hybridization with the mirror symmetry 
breaking below $T_c$ but without $e$-$h$ 
correlation.\cite{subedi}

Following the above development, it is important 
to carefully evaluate effects of both the $e$-$h$ 
correlation and $e$-$l$ coupling in Ta$_2$NiSe$_5$.  
An important aspect is the response of 
Ta$_2$NiSe$_5$ to changes in the ratio 
$E_b/E_g$, since the physical properties of EI 
should sensitively depend on $E_b/E_g$.   
In this regard, photo-excited experiments, which 
are probed by either ARPES\cite{mor1,okazaki1,mizokawa,tang,
okazaki2,saha} or laser,\cite{weudehausen,mor2,okamoto} 
have the advantage of being able to tune the carrier 
density, and hence to control $E_b$ through the Coulomb 
screening of $e$-$h$ attraction.  
In fact, photo-induced gap closing and semimetallic 
states in Ta$_2$NiSe$_5$ have been observed and analyzed.
\cite{mor1,okazaki1,mizokawa,tang,okazaki2,saha,okamoto}  
Another effective way of tuning an EI state is the 
application of external pressure 
($P$),\cite{lu,sood,sawa,matsubayashi} since an applied 
pressure usually broadens the conduction and 
valence bands, and increases (reduces) their 
overlap (separation).  
Detailed high-$P$ studies\cite{sawa,matsubayashi} 
have been performed on the crystal structure, 
resistivity ($\rho$), Hall coefficient, and 
$\sigma(\omega)$ of Ta$_2$NiSe$_5$.  
They revealed a rich $P$-$T$ phase diagram of 
Ta$_2$NiSe$_5$ as depicted in Fig.~2.  
\begin{figure}[t]
\begin{center}
\includegraphics[width=0.44\textwidth]{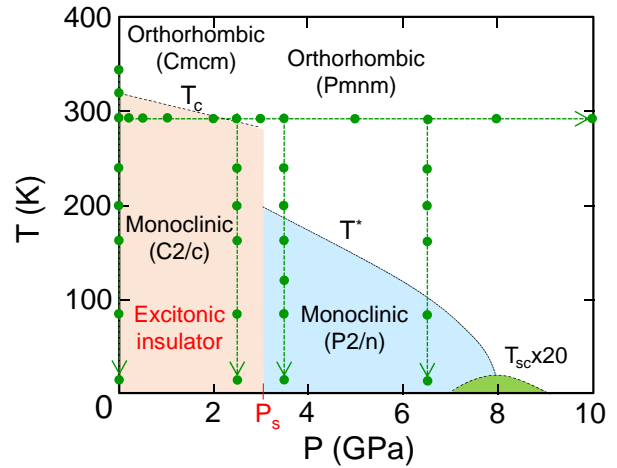}
\caption{Temperature ($T$)-pressure ($P$) phase 
diagram of Ta$_2$NiSe$_5$ reproduced from 
Matsubayashi {\it et al.}\cite{matsubayashi}   
$T_c$, $T^\ast$, and $P_s$ indicate the structural 
transition temperatures and pressure, and $T_{sc}$ 
the superconducting transition temperature.
The green arrows and dots indicate the ($P,T$) 
paths and points where $\sigma(\omega)$ was 
measured in this work.}
\end{center}
\end{figure}
Above the first order structural transition 
at $P_s \simeq$ 3~GPa, the $ac$ plane becomes 
flatter with a less rippling of $ac$ plane 
along the $c$ axis.\cite{sawa,matsubayashi}  
$\rho(T)$ is reduced with increasing $P$, showing 
a crossover from semiconducting $T$ dependence 
below $P_s$ to a completely metallic one well above 
$P_s$, with quite large and dramatic change of 
$\rho(T)$ around $P_s$.  
A superconductivity with $T_c$=2~K appears 
at about 8~GPa.    
Above $P_s$, from $\rho(T)$ and $\sigma(\omega)$ 
data, it has been shown that Ta$_2$NiSe$_5$ is 
a semimetal with a partial energy gap, 
in contrast to the full energy gap below $P_s$.  
It has been suggested that the electronic transition 
at $T^\ast$ to the low-$T$ semimetallic state is 
caused by $e$-$l$ coupling, while that at $T_c$ 
below $P_s$ is caused by both excitonic 
correlation and $e$-$l$ coupling.\cite{matsubayashi}

In this paper, we present a full account of the 
$\sigma(\omega)$ and reflectance data of Ta$_2$NiSe$_5$ 
at high $P$ and low $T$, which have been obtained 
at the ($P$, $T$) points indicated in Fig.~2.     
The data reveal how the well-developed energy gap 
and excitonic peak at $P$=0, which have been 
taken as evidence for an EI, are suppressed as the 
applied pressure reduces the excitonic correlation.

\section{Experimental}
The samples of Ta$_2$NiSe$_5$ used were single 
crystals grown with the chemical vapor transport 
method as reported previously.\cite{lu}  
The reflectance spectrum at zero external pressure 
[$R_0(\omega)$] was measured at 0.015-4.5~eV photon energy 
range\cite{okamura-chapter} on an as-grown, specular surfaces.  
$\sigma(\omega)$ was derived from the measured $R_0(\omega)$ 
using the standard Kramers-Kronig (KK) analysis.\cite{dressel}  
Below the measured energy range, $R_0(\omega)$ was 
extrapolated by the Hagen-Rubens function or a constant, 
depending on the data.\cite{dressel}   
Above the measured range, it was extrapolated by a 
function of the form $\omega^{-d}$.\cite{gaisou}   
$T$ dependence of $R_0(\omega)$ were measured 
below 1.4~eV, and above 1.4~eV the $R_0(\omega)$ 
at 295~K was connected.   
Reflectance spectra at high $P$ were measured 
using a diamond anvil cell (DAC).\cite{pressure-review}  
Type IIa diamond anvils with a 0.8~mm culet diameter 
and a stainless steel gasket were used to seal the 
sample with KBr as the pressure transmitting medium.   
An as-grown surface of a sample was directly attached 
on the culet surface of the diamond anvil, and the 
reflectance at the sample/diamond interface 
[$R_d(\omega)$] was measured.    A gold film was also 
sealed with the sample as the reference of reflectance.  
The pressure in the DAC was measured with the ruby 
fluorescence method.\cite{NBS}     
$R_d(\omega)$ at high $P$ and low $T$ were measured 
at photon energies between 0.025 and 1.1~eV, 
using synchrotron radiation as a bright infrared 
source\cite{JPSJ-review} at the beamline BL43IR of 
SPring-8.\cite{micro}   The measured $R_d(\omega)$ 
spectra below 1.1~eV were connected to the 
$R_d(\omega)$ above 1.1~eV given by the KK analysis 
of measured $R_0(\omega)$ at 295~K and the refractive 
index of diamond ($n_d$=2.4).  
Then the connected $R_d(\omega)$ spectra were used 
to obtain $\sigma(\omega)$ by a modified KK 
analysis.\cite{okamura-kk} 
More details of the high pressure infrared experiments 
are reported elsewhere.\cite{pressure-review}

\section{Results and Discussions}
%
In the crystal structure of Ta$_2$NiSe$_5$, the (-Ta-Se-) 
and (-Ni-Se-) chains extend along the $a$ axis in the 
$a$-$c$ layers stacked along the $b$ axis.\cite{sunshine,salvo}   
Therefore, characteristic optical responses 
of EI are expected with $E \parallel a$ polarization 
where $E$ is the electric field of the incident light.   
$R_0(\omega)$ and $\sigma(\omega)$ spectra of Ta$_2$NiSe$_5$ 
measured at ambient pressure ($P$=0) are indicated in 
Figs.~3(a) and 3(b), respectively.
%
\begin{figure}[t]
\begin{center}
\includegraphics[width=0.425\textwidth]{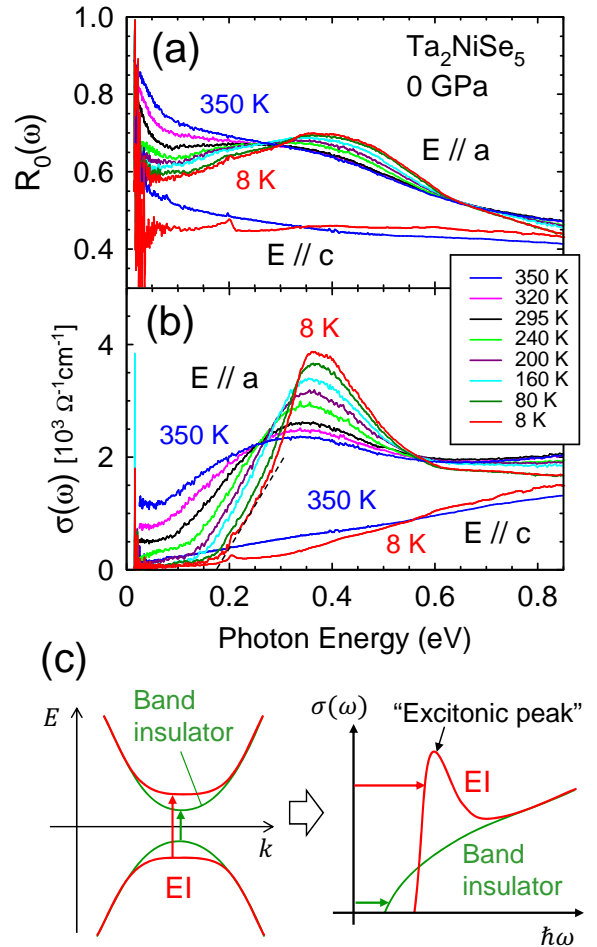}
\caption{(a) Optical reflectance [$R_0(\omega)$] and (b) 
conductivity [$\sigma(\omega)$] of Ta$_2$NiSe$_5$ at 
different temperatures and at ambient pressure.  
$E \parallel a$ and $E \parallel c$ denote incident 
light polarized along $a$ and $c$ axes, respectively.  
The broken line in (b) is a guide to the eye, 
indicating the onset at 0.17~eV.  
(c) Schematic diagrams indicating the optical 
transitions in an EI (red curves) and a band 
insulator (green curves).  
}
\end{center}
\end{figure}
%
In the $E \parallel a$ data, with cooling, $R_0(\omega)$ 
exhibits large $T$ dependences.  $\sigma(\omega)$ develops 
a clear energy gap with the onset at about 0.17~eV, as 
indicated by the broken line in Fig.~3(b), and a pronounced 
peak centered at $\simeq$ 0.38~eV.\cite{footnote1}   
The sharp spikes below 0.05~eV are due to optical 
phonons and the periodic oscillations below 0.1~eV 
are due to interference caused by internal reflections 
from the rear surface of the sample.   
In contrast, in the $E \parallel c$ data, both $R_0(\omega)$ 
and $\sigma(\omega)$ are much lower than those in the 
$E \parallel a$ data, and their $T$ dependence 
is much smaller.  At 8~K, the onset of $\sigma(\omega)$ appears 
to be located around 0.3~eV, with only a gradual 
increase of $\sigma(\omega)$ above the onset.    
These results are very similar to previously 
reported $P$=0 data in the literature.\cite{lu,boris,seo}  
The development of the energy gap and the pronounced 
peak with $E \parallel a$ have been 
interpreted as the result of EI state formation below $T_c$.  
The peak at 0.38~eV is an excitonic peak, as already 
discussed in Introduction, and its origin can be basically 
understood in terms of a band flattening as 
illustrated in Fig.~3(c).  
%
%
\begin{figure*}
\centering
\includegraphics[width=0.94\textwidth,clip]{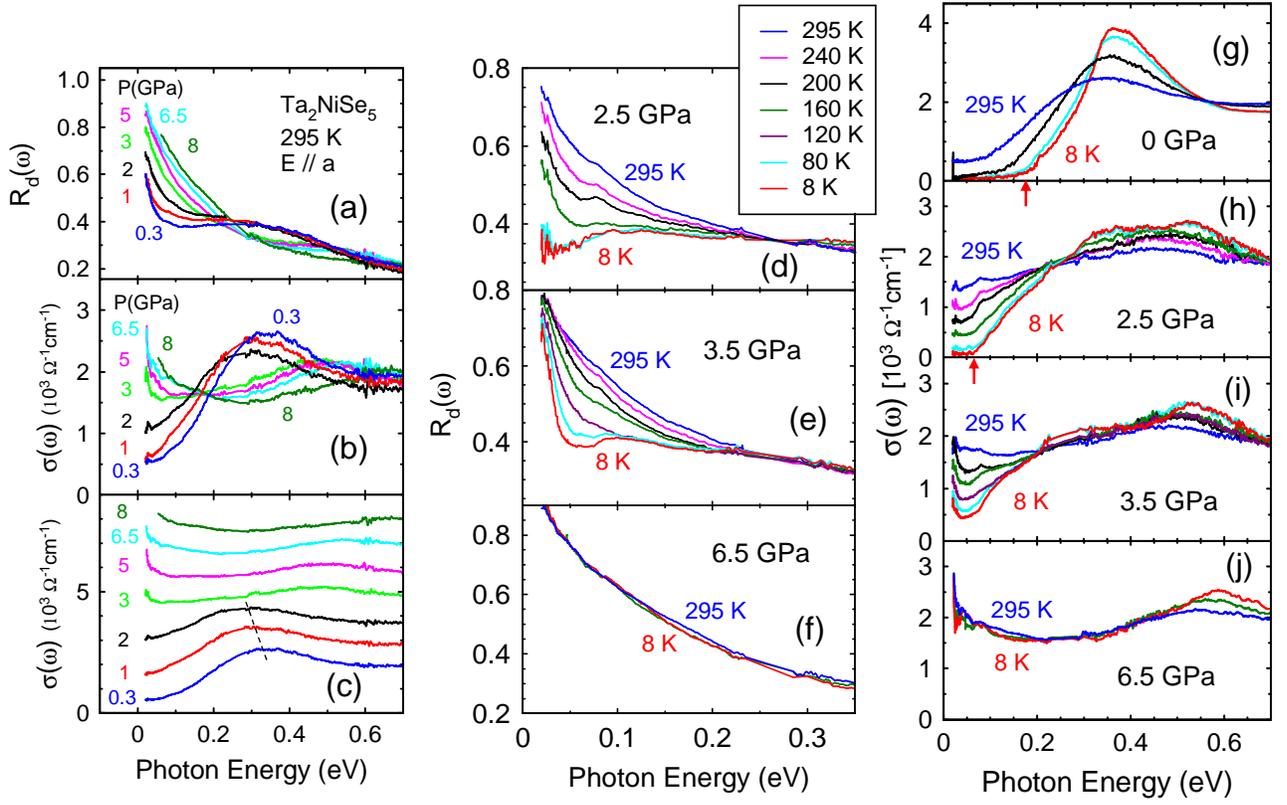}
\caption{(a) Optical reflectance [$R_d(\omega)$] and (b)(c) 
conductivity [$\sigma(\omega)$] of Ta$_2$NiSe$_5$ at 
295~K under high pressure ($P$).  
$R_d(\omega)$ in the 0.235-0.285~eV range has been interpolated 
by a straight line since it could not be measured well due 
to strong absorption by diamond.\cite{pressure-review}   
In (c), the same spectra as those in (b) are vertically offset, 
and the broken line indicates the shift and suppression of 
the excitonic peak.  
(d)-(f) $R_d(\omega)$ of Ta$_2$NiSe$_5$ measured at 
$P$=2.5, 3.5, and 6.5~GPa and at different temperatures.  
(g)-(j) $\sigma(\omega)$ of Ta$_2$NiSe$_5$ at $P$=0, 
2.5, 3.5, and 6.5~GPa and at low temperatures. The red 
vertical arrows indicate the onset of $\sigma(\omega)$ 
discussed in the text. The same temperature labels 
(the colors of the curves) apply in (d)-(f) and (g)-(j).  
}
\end{figure*}
In a conventional insulator, the optical excitation 
between the band edges lead to the increase of 
$\sigma(\omega)$ above the band gap, as indicated by 
the green curves in Fig.~3(c).  The spectrum in this 
case is actually similar to that measured with 
$E \parallel c$ in Fig.~3(b).  In the case of EI, 
in contrast, the band edges are flattened by 
excitonic correlation.  This results in a larger 
density of states  (DOS) at the band edges, which 
should cause an enhancement in $\sigma(\omega)$ as 
shown by the red curves in Fig.~3(c).  
The spectrum in this case is very similar to 
that observed with $E \parallel a$ in Fig.~3(b).    
The $T$ evolution of $\sigma(\omega)$ in Ta$_2$NiSe$_5$ 
has also been discussed in terms of exciton-phonon 
Fano effect\cite{boris} and exciton superfluid 
formation.\cite{seo}   
The shifts of the onset and the transfer of spectral 
weight in $\sigma(\omega)$ will be discussed later.

Pressure evolution of $R_d(\omega)$ and $\sigma(\omega)$ 
at 295~K with $E \parallel a$ is shown in Figs.~4(a)-4(c).   
Upon applying pressure, both $R_d(\omega)$ and $\sigma(\omega)$ 
below 0.2~eV exhibit large increases.  In particular, 
$\sigma(\omega)$ below $\sim$ 0.2~eV, which is reduced by 
the energy gap at $P$=0, rapidly increases with $P$.  
In addition, the excitonic peak is broadened and 
red-shifted with $P$, and is almost suppressed 
at 3~GPa and above.  
This is more clearly seen in Fig.~4(c) where the spectra 
are vertically offset.  
The large spectral change at around 3~GPa is almost 
coincident with the structural transition at 
$P_s$ (Fig.~2).  
Since $P_s$ is believed to be the boundary of EI 
phase,\cite{matsubayashi} it is reasonable that the 
excitonic peak is suppressed above $P_s$. 
The broadening and red shift of the excitonic 
peak below $P_s$ probably indicate the weakening 
of excitonic correlation even within the 
EI phase below $P_s$.  
As discussed in Appendix, Raman spectra of 
Ta$_2$NiSe$_5$ at high $P$ were also measured, 
using the same DAC and sample condition.   
The Raman spectra (Fig.~7) exhibit a sudden shift 
and disappearance of phonon peaks around 3~GPa, 
which should result from the first order structural 
transition at $P_s$.  
This Raman result further supports that the observed 
spectral changes in $\sigma(\omega)$ indeed result 
from the structural transition at $P_s$.  
With increasing $P$ above 3~GPa, $\sigma(\omega)$ 
further increases and shows metallic characteristics 
with a rising (Drude) component toward zero energy.

Temperature dependent $R_d(\omega)$ spectra at 
$P$=2.5, 3.5 and 6.5~GPa are indicated in 
Figs.~4(d)-(f), and the corresponding $\sigma(\omega)$ 
spectra in Figs.~4(h)-4(j).  
First, note that the large $T$ dependences of 
$R_d(\omega)$ and $\sigma(\omega)$ observed at $P$=0 
are progressively reduced with increasing $P$.   
Accordingly, both the energy gap and excitonic peak 
at low $T$ are progressively suppressed with $P$.  
At 2.5~GPa, $\sigma(\omega)$ develops a clear 
energy gap with cooling [Fig.~4(h)], but the onset 
of $\sigma(\omega)$ at 8~K is located at 0.062~eV, 
which is much smaller than that at $P$=0 [Fig.~4(g)].  
Namely, a clear energy gap still exists at $P$=2.5~GPa, 
but its width is significantly reduced from 
that at $P$=0.    
At 3.5~GPa, $\sigma(\omega)$ below 0.2~eV is still 
strongly reduced with cooling, but unlike those at 
$P$=0 and 2.5~GPa, it is not completely depleted even 
at 8~K.   A rising component toward zero energy is 
observed 8~K, which should be a Drude component 
due to free carriers.  
The thermal energy $k_{\rm B} T$, where $k_{\rm B}$ 
is the Boltzmann constant, is about 0.7~meV at 8~K, 
while the tail of the Drude component at 8~K extends 
to much higher energy, $\sim$ 50~meV.  
Therefore, the electronic state at 3.5~GPa 
and 8~K is difficult to understand as a semiconductor 
with thermally excited free carriers.   Instead, it 
should be a semimetal with a band overlap of the order 
of 50~meV.  Accordingly, the energy gap at 3.5~GPa 
should be a partial one, which is open only in certain 
portions of the Brillouin zone.\cite{matsubayashi}  
At 6.5~GPa, the spectra show only minor $T$ dependences, 
and $\sigma(\omega)$ below 0.4~eV is almost unchanged 
from 295 to 8~K.   Clearly, at 6.5~GPa there is no 
energy gap in Ta$_2$NiSe$_5$ and its electronic 
structure is quite metallic.

To examine more closely the variations of energy 
gap with $P$ and $T$, the low-energy portion of 
$\sigma(\omega)$ at $P$=0 and 2.5~GPa are 
displayed in Figs.~5(a) and 5(b), respectively.   
\begin{figure}
\begin{center}
\includegraphics[width=0.45\textwidth]{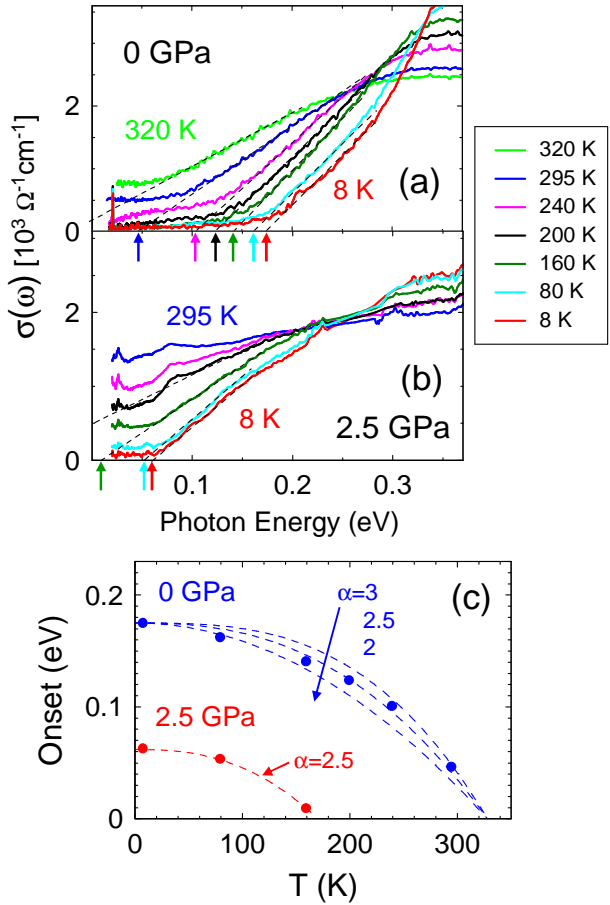}
\caption{Low-energy portion of $\sigma(\omega)$ in 
$E \parallel a$ polarization at 
(a) $P$=0 and (b) $P$=2.5~GPa.   
The broken lines indicate extrapolations 
to the linear-in-energy portion of the spectra, and 
the vertical arrows indicate the onset 
given by the zero crossing.   
(c) The onset energy in (a) and (b) plotted as 
a function of temperature ($T$).   The broken curves 
indicate $\Delta(T)=\Delta(0) \{ 1-(T/T_c)^\alpha \}$ 
with the indicated values of $\alpha$.  
$\Delta(0)$=0.175~eV and $T_c$=328~K are used for 
$P$=0, and $\Delta(0)$=0.062~eV and $T_c$=170~K 
for $P$=2.5~GPa.  
}
\end{center}
\end{figure}
The broken lines are extrapolations 
to the linear-in-energy portion in $\sigma(\omega)$.  
Their zero crossings, marked with the vertical arrows, 
indicate the onset of $\sigma(\omega)$.   
Here we regard the onset energy as the gap width 
$\Delta$, following a previous $\sigma(\omega)$ work 
on Ta$_2$NiSe$_5$ (Ref.~[\onlinecite{seo}]) and other 
optical studies on strongly correlated electron 
systems.\cite{basov}   
At $P$=0 [Fig.~5(a)], the extrapolation crosses 
zero at 295~K and below, but not at 320~K.  Namely, 
the gap in $\sigma(\omega)$ becomes fully open at 
295~K and below, which 
is reasonable since the gap formation starts below 
$T_c$=328~K and the gap is not well developed yet 
at 320~K.  The $T$ dependence of the onset energy 
is plotted in Fig.~5(c).  The result is similar 
to that previously reported by Ref.~[\onlinecite{seo}], 
which reported a $T$ dependence of the form 
$\Delta(T)=\Delta(0)\{1-(T/T_c)^\alpha \}$ 
with $\alpha$=2.   
For comparison, in Fig.~5(c) we also plot this 
function with $T_c$=328~K and $\alpha$=2, 2.5, and 3.  
It is seen that the present data is close to 
$\alpha$=2.5, but cannot be well described 
by a single exponent.  
At 2.5~GPa [Fig.~5(b)], the extrapolation crosses 
zero at 160~K and below, but not at 200~K.  Namely, 
the gap becomes fully open at 160~K and below, which 
is much lower than $T_c \simeq$ 290~K at 2.5~GPa.   
This is in contrast to the $P$=0 case  above, where 
the gap in $\sigma(\omega)$ is already open 
at 295~K which is below but still close to 
$T_c$=328~K.  
Namely, it seems that the $P$ dependence of the energy 
gap in $\sigma(\omega)$ is different from that of $T_c$.  
The $T$ dependence of the onset at 2.5~GPa is also 
plotted in Fig.~5(c).  It is similar to that at $P$=0, 
but to fit with the form 
$\Delta(T)=\Delta(0)\{1-(T/T_c)^\alpha \}$, 
$T_c \simeq$ 170~K and $\alpha \simeq$ 2.5 are 
required as shown by the broken curve in Fig.~5(c).   
Note that the use of this functional form is only 
phenomenological, and not based on a microscopic 
model for EI.  
The much stronger suppression of energy gap by 
pressure than that of $T_c$ will be discussed 
again later.

To analyze the spectral weight (SW) transfer in 
$\sigma(\omega)$ with $P$ and $T$, the 
effective electron number $N^\ast(\omega)$ per 
formula unit (fu) of Ta$_2$NiSe$_5$, 
expressed as \cite{dressel} 
\begin{equation}
N^\ast(\omega)=\frac{n}{m^\ast}(\omega)
=\frac{2 m_0}{\pi e^2 N_0}\int_{0}^{\omega}
\sigma(\omega^\prime)d\omega^\prime,
\end{equation}
is plotted in Fig.~6.  
%
\begin{figure}[t]
\begin{center}
\includegraphics[width=0.35\textwidth]{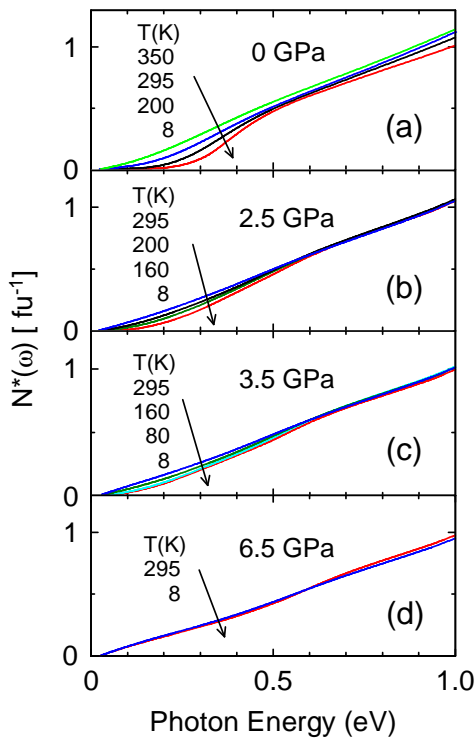}
\caption{Effective electron number $N^\ast(\omega)$ per 
formula unit (fu) of Ta$_2$NiSe$_5$, calculated with 
Eq.~(1) and the measured $\sigma(\omega)$ spectra at 
different pressures and temperatures.  
}
\end{center}
\end{figure}
Here, $n$, $m^\ast$ and $e$ are the electron density, 
effective mass in units of the rest mass $m_0$, 
and the elementary charge.    
$N_0$ is the number of fu's per unit volume, which 
was calculated using the high $P$ crystal structure 
data.\cite{sawa,footnote4}   
$N^\ast(\omega)$ can be regarded as the number of 
electrons that contribute to $\sigma(\omega)$ between 0 
and $\omega$, therefore it directly reflects the SW 
transfer up to $\omega$.      
At all $P$ and $T$ in Fig.~6, $N^\ast(\omega)$ at 1~eV 
is of the order of 1~fu$^{-1}$, which is reasonable 
compared with the total DOS within 1~eV from $E_{\rm F}$ 
given by band calculations.\cite{matsubayashi}  
At $P$=0, $N^\ast(\omega)$ varies with $T$ over a wide 
energy range.  The $N^\ast(\omega)$ spectra at different 
temperatures do not converge even at 1~eV, showing that 
the electronic structure reconstruction in the EI phase 
occurs over a wider energy scale than the gap width 
(0.17~eV) and the excitonic peak energy (0.4~eV).   
At 2.5 and 3.5~GPa, however, $N^\ast(\omega)$ varies 
with $T$ only up to about 0.6~eV, indicating that the 
SW transfer is complete within 0.6~eV.   At 6.5~GPa, 
$N^\ast(\omega)$ is almost independent of $T$.  
These results show that both the amount and energy 
range of the SW transfer are progressively suppressed 
by $P$, which is closely related with the 
suppression of energy gap.

As discussed in Introduction, it has been pointed out 
that the gap formation in Ta$_2$NiSe$_5$ below $P_s$ 
may be partly due to $e$-$l$ coupling, rather than 
being solely due to $e$-$h$ (excitonic) correlation.  
If their relative contributions do not change with 
$P$, and if both the structural transition at $T_c$ 
and the gap opening below $T_c$ are driven by a 
common mechanism, then one would naively expect that 
$\Delta$ and $T_c$ should have similar $P$ dependences.  
This is in contrast to the present results, in 
which the reduction of $\Delta$ by pressure is 
much greater than that of $T_c$ 
(from $\Delta$=0.17~eV and $T_c$=328~K at $P$=0 
to $\Delta$=0.062~eV and $T_c \simeq$ 290~K at 
2.5~GPa).   
Namely, in terms of the $P$ dependence, $\Delta$ 
is not scaled with $T_c$. 
Interestingly, a pressure decoupling between 
electronic and structural transitions has been 
observed in BaFe$_2$As$_2$,\cite{Ba122} which 
is another pressure-induced superconductor.  
The present results for Ta$_2$NiSe$_5$ may suggest 
that, somehow, the $e$-$h$ correlation is more 
responsible for the gap development below $T_c$, 
while the $e$-$l$ coupling is more responsible 
for the structural transition at $T_c$.   
This would naturally lead to a stronger 
suppression of $\Delta$ than $T_c$ by pressure, 
since the $e$-$h$ correlation is likely more 
reduced by pressure than the $e$-$l$ coupling 
as long as the structure does not change.  
An understanding of these $P$ dependences 
based on a specific microscopic model is 
beyond the scope of the present study.   
It is hoped that effects of both 
the $e$-$h$ correlation and 
the $e$-$l$ coupling with symmetry 
breaking-induced hybridization 
are both included in a microscopic model 
to understand the pressure evolution of 
electronic states in Ta$_2$NiSe$_5$.

\section{Summary}
$\sigma(\omega)$ spectra of Ta$_2$NiSe$_5$, 
a strong candidate for an EI, have been measured 
at high pressures and at low temperatures.  
A clear energy gap of about 0.17~eV and a marked 
excitonic peak at 0.38~eV at $P$=0 are 
suppressed with increasing $P$, which is 
especially marked across the structural 
transition at $P_s$=3~GPa.   
At 2.5~GPa ($< P_s$), the gap is still 
clearly open with a reduced width of 0.062~eV, 
but at 3.5~GPa ($>P_s$), the gap is 
partially filled, indicating that the 
ground state above $P_s$ is semimetallic.   
At 6.5~GPa, $\sigma(\omega)$ shows very metallic 
characteristics with no energy gap.   
These results have been discussed in terms of 
a progressive weakening of excitonic correlation 
with pressure.  
Also, it is noted that the reduction of energy 
gap with pressure is much greater than that of $T_c$.  
This result may suggest that the relative importance 
of $e$-$h$ correlation and $e$-$l$ coupling 
in the gap formation changes with pressure.

\begin{acknowledgments}
We would like to thank Prof. K. Matsubayashi, 
Prof. Y. Ohta, and Prof. K. Sugimoto for useful 
discussions.  
Experiments at SPring-8 were performed under the 
approval by JASRI (Proposal numbers 
2013A1085, 
2015B1698, 2016A1166, 
2017A1163, 
2017A1164.) 
This work was supported by KAKENHI 
(21102512, 23540409, 26400358, 
19H00659, 
17H06456)
\end{acknowledgments}

\appendix* \section{Raman spectra at high pressures}
Raman scattering spectra of Ta$_2$NiSe$_5$ 
measured at high pressures and at 295~K are 
shown in Fig.~7.  
%
\begin{figure}[b]
\begin{center}
\includegraphics[width=0.4\textwidth]{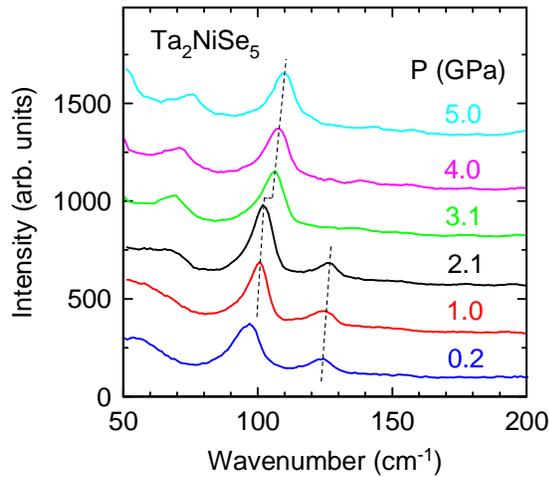}
\caption{Raman spectra of Ta$_2$NiSe$_5$ measured 
at high pressures ($P$) and at 295~K.  
The spectra are vertically offset for clarity.  
The excitation was made at 532~nm with 
$E \parallel a$ polarization.   The broken 
lines are guide to the eye, emphasizing the 
sudden changes at around 3~GPa.   
}
\end{center}
\end{figure}
The measurement was made under a back scattering 
geometry on an as-grown sample surface containing 
the $ac$ plane.   A standard micro-Raman apparatus 
was used with a 532~nm laser source polarized 
along the $a$ axis.  
High pressures were applied using the same DAC 
as that used for the $\sigma(\omega)$ studies, 
with exactly the same procedure for sample loading.  
In Fig.~7, the Raman peaks are due to optical 
phonons.\cite{yan,raman-MPI,sood} 
As indicated by the broken lines in Fig.~7, the 
peak near 100~cm$^{-1}$ is blue shifted and the 
peak near 125~cm$^{-1}$ disappears in the spectra 
at 3.1~GPa and above.   
These sudden changes should result from the 
first-order structural phase transition at $P_s$ 
(Fig.~2), where the rippling of the $ac$ plane 
along $c$ axis is reduced 
above $P_s$.\cite{sawa,matsubayashi}  



\begin{thebibliography}{99}
\bibitem{gaisou}
Above the measured range, $R_0(\omega)$ was first 
extrapolated with a function of the form 
$\omega^{-0.7}$ up to 15~eV, then with 
$\omega^{-4}$ up to 35~eV.  The choice of these 
functions is 
certainly arbitrary, but they were chosen so that 
the $\sigma(\omega)$ obtained with the KK analysis 
of the extrapolated $R_0(\omega)$ was consistent 
with the previously reported $\sigma(\omega)$ 
measured with the ellipsometry.\cite{boris}
\bibitem{footnote1}
The present data at $P$=0 have been 
measured on more recent and larger samples than 
those used in Ref.~[\onlinecite{matsubayashi}].   
Although the present $R_0(\omega)$ spectra in the 
infrared range [Fig.~3(a)] are similar to those 
used in Ref.~[\onlinecite{matsubayashi}], the present 
$\sigma(\omega)$ spectra [Fig.~3(b)] are about 30~\% 
higher and their onset at low $T$ is much clearer than 
those in Ref.~[\onlinecite{matsubayashi}].  
These differences have mainly resulted from a much higher 
$R_0(\omega)$ in the visible-UV range in the present 
data and the high-energy extrapolations used.  
Note that a change in $R_0(\omega)$ at a higher energy 
range may cause a change in $\sigma(\omega)$ at a lower 
energy range through the KK relation.\cite{dressel}
\bibitem{footnote4}
The volume per fu at room temperature and 
$P$=0, 2.5, 3.5, and 6.5 GPa have been 
estimated to be 173, 165, 159, and 150~$\AA^3$, 
respectively, from the crystal structure data 
at high $P$.\cite{sawa}  
In the calculation of $N^\ast(\omega)$, the 
volume contraction with cooling has been 
neglected, since it is much smaller than 
that with increasing $P$.
\end{thebibliography}
\end{document}